\documentclass[11pt,citesort]{article}
\usepackage{color}
\usepackage[dvips]{graphicx}

\usepackage{amsmath,amssymb}

\newcommand{\bel}[1]{\begin{equation}\label{#1}}
\newcommand\be{\begin{equation}}
\newcommand\ee{\end{equation}}
\newcommand\bea{\begin{eqnarray}}
\newcommand\eea{\end{eqnarray}}
\newcommand{\bdm}{\begin{displaymath}}
\newcommand{\edm}{\end{displaymath}}
\newcommand\nn{ \nonumber\\}

\newcommand{\dd}[1]{\partial_{#1}}
\def\dd{\partial}
\renewcommand{\>}{\rangle}

\makeatletter
\@addtoreset{equation}{section}
\makeatother

\title{On the Eikonal Approximation in AdS Space}
  
\author{ Richard  C. Brower\footnote{Physics Department,
Boston University, Boston MA 02215},\
Matthew J. Strassler\footnote{Department of Physics, University of Washington, Seattle,
  WA 98195} \
and  Chung-I Tan\footnote{Physics Department, Brown University,
Providence, RI 02912}\\ 
\\
E-mails: \\
brower@bu.edu, strassler@rutgers.edu,  tan@brown.het.edu
}

\begin{document}
 
\maketitle

\begin{abstract}
We explore the eikonal approximation to graviton exchange in $AdS_5$
space, as relevant to scattering in gauge theories.  We restrict
ourselves to the regime where conformal invariance of the dual gauge
theory holds, and to large 't Hooft coupling where the computation
involves pure gravity.  We give a heuristic
argument, a direct loop computation, and a shock wave derivation.  The
scalar propagator in $AdS_3$ plays a key role, indicating that even at
strong coupling, two-dimensional conformal invariance 
controls high-energy
four-dimensional gauge-theory scattering.
\end{abstract}

\section{Introduction}

There has been much interest in high-energy scattering, and the
eikonal approximation in particular, in the contexts of gauge
theories, string theories, and the duality which relates them.
Relevant recent papers  
include \cite{Veneziano:2004er,BPST,Cornalba:2006xk,Cornalba:2006xm,Giddings:2007bw,Alday:2007hr}.
In this note, we obtain the eikonal approximation to scattering in
$AdS_5$.  This is relevant for high-energy limits of Green functions in
four-dimensional conformal field theory, and plays a partial role
in  dual descriptions of high-energy hadron scattering
and small-$x$ deep-inelastic scattering in nonconformal quantum
field theories.  Applications will be addressed elsewhere; here
we present some basic results.\footnote{Work with overlapping results
appeared \cite{Cor} as this paper was in preparation.}

The string-dual description of scattering at very high energies and
low momentum transfer in large-$N$, large-$\lambda$ gauge theories
(here $\lambda\propto g^2N$ is the 't Hooft coupling) was considered
in~\cite{BPST}, building on work of \cite{Polchinski:2001tt,DIS,Brower:2002er}. 
The dual string theory has strings propagating on a
space which is asymptotically $AdS_5\times X$ ($X$ a compact
five-dimensional space of little consequence here) with metric
approaching
\bel{AdSXmetric}
ds^2 \approx {R^2\over z^2}(\eta_{\mu\nu}dx^\mu dx^\nu+dz^2) + ds_X^2  \ 
\ee
as $z\to 0$.  The metric is strongly corrected for $z$ near
$z_{max}\sim 1/\Lambda$, where $\Lambda$ is the confinement scale.

The form of a two-to-two hadron scattering amplitude in the gauge
theory, at leading order in $1/N$ and $s$ large compared to the
confinement scale and to $t$, was shown in \cite{BPST}, following
\cite{DIS}, to be proportional to a Pomeron propagator or ``kernel''
${\cal K}_0$.
\begin{equation}\label{ourBFKL}
{\rm Im}\left[{\cal M}_{2\to2}(s,t)\right] \propto  {
\int dz \sqrt{g(z)}\,{dz'} \sqrt{g(z')}\, }
\Phi_3(z)\Phi_4(z) 
\left[{\tilde s^2 
\left({ zz' \over R^2}\right)^{2}}{\cal K}_0(s,t,z,z')\right] 
\Phi_1(z') \Phi_2(z') \ .
\end{equation}
Here the $\Phi_i$ are wave functions for states dual to gauge-theory hadrons,
and the variable $\tilde s$ represents the square 
of the proper\footnote{{\it i.e.,} measured
by a local observer in the bulk,} center-of-mass energy.
\be
\tilde s\equiv {s \over \sqrt{g_{+-}(z)g_{-+}(z')}}={zz' s\over R^2} \ .
\ee
(Note $\tilde s$ is slightly ambiguous, 
though generally the ambiguity is subleading in the Regge limit;
we have taken $\tilde s$ to
have a symmetric dependence on $z$ and $z'$, as was done in \cite{BPST}.)  
An analogous formula for deep-inelastic scattering at small-$x$ was
obtained earlier in \cite{DIS}.  One may obtain similar formulas from
four-point functions, as in \cite{Freedman}, by taking high-energy
limits of their (suitably regulated) Fourier transforms.  In each of
these cases, the amplitude resembles (\ref{ourBFKL}); there are four
external functions $\Phi_i(z)$ which are normalizable or
nonnormalizable modes of fields in the bulk, combined with a kernel
${\cal K}_0$.

In a conformal field theory, or in a confining theory with $t\ll
-\Lambda^2$, the kernel takes the form 
\bel{thekernel} {\cal
  K}_0(s,t,z,z') = {2\over \pi^2}s^{J_0-2} \int_0^\infty\ d\nu\, \nu\,
\sinh \pi\nu \ K_{i\nu} (z |t|^{1/2}) K_{-i\nu} (z' |t|^{1/2})
e^{-\nu^2 \tau}\ .  
\ee 
where $\tau = {1\over 2\sqrt{\lambda}}\ln
(s/s_0)$.  This is similar in form to the BFKL kernel
\cite{BFKL1,BFKL2,BFKL3}, matching its form precisely (but not its
coefficients, which are $\lambda$-dependent) for $t=0$, and having
similar form for large $t$.  Like the BFKL kernel, its analytic
structure involves a cut\footnote{In both cases, the cut becomes a discrete
  and dense set of poles when the coupling runs slowly.}  in the
$J$-plane extending along the negative real axis starting from $J =
J_0$, 
\bel{j0} J_0=2-{2\over\sqrt{\lambda}} +O(\lambda^{-1}) . 
   \ee

As $\lambda\to\infty$, $s$ fixed, the scattering amplitude is
described by a pure graviton exchange, with $J_0\to 2$.  
Define the amputated four-point amplitude
${\cal A}^{(1)}$ for a single $t$-channel Pomeron exchange by
\bea
{\cal A}^{(1)}(s,x^\perp,z;y^\perp,w;x'^\perp,z';y^\perp,w) &=& 
{\cal M}^{(1)}(s,x^\perp,z;x'^\perp,z') \left[g(z)g(z')\right]^{-1/2}
\ \ \ \ \ \ \ 
 \\ &  &  \ \times
\delta^2(x^\perp-y^\perp)\delta(z-w)\delta^2(x'^\perp-y'^\perp)\delta(z'-w')
\nonumber
\eea
In
the limit $\lambda$ large, $s^2{\cal K}_0$, up to some metric factors and delta
functions, is just the imaginary part of ${\cal M}^{(1)}$, 
\be
{1\over 2i}{\rm Disc}\left[{\cal M}^{(1)}(s,x^\perp,z;x'^\perp,z')\right] 
\sim  { \tilde s^2 
\int d^2x^\perp e^{iq_\perp \cdot x^\perp}
\left({ zz' \over R^2}\right)^{2}} {\cal K}_0(s,q_\perp^2,z,z') 
\ee
The full amplitude can be constructed from the discontinuity and
crossing-symmetry as usual through methods
of analyticity.  A short calculation using Eqs.~(\ref{ourBFKL}) and
(\ref{thekernel}) reveals, as $\lambda \rightarrow \infty$, 
\be
{\cal M}^{(1)}(s,x^\perp,z;x'^\perp,z') =
{\kappa_5^2\over  R} \tilde s^{2} \left(zz'\over R^2\right)
G_3(x^\perp,z;x'^\perp,z') \label{eq:gravitontree}
\ee
where $\kappa_5$ is the gravitational coupling in $AdS_5$, 
$G_3$ is the dimensionless scalar propagator for a particle of mass
$\sqrt{3}/R$ in an Euclidean $AdS_3$ space of curvature radius $R$, i.e., a propagator over a three dimensional hyperboloid,
\bel{Gfreedman}
G_3(x^\perp,z;x'^\perp,z')= G_3(u) 
=\frac{1}{ 4\pi  }\frac{ 1}{\left[1+u + \sqrt {u(2+u)}\right]^2 \sqrt{u(2+u)}}
\ee
and
\bel{udef}
u = {\delta_{ij}(x^\perp-x'^\perp)^i(x^\perp-x'^\perp)^j
+ (z-z')^2\over 2 z z'}
\ee 
(with $i,j=1,2$)  is the chordal distance on the $AdS_3$ transverse to the 
momentum direction of the scattering particles.  
We will explain in the next section why this result should be expected. 

Since the amplitude grows faster than $s$, it violates unitarity at
large $s$, and a resummation of higher-loop amplitudes is required to
obtain sensible physics.  In certain restricted regions of $z$ and
$z'$, this resummation can be done via the eikonal approximation.
Note however that a complete field-theory computation, which requires
integrating over $z$ and $z'$, typically is not possible within the
bulk eikonal approximation.  With this caveat, we proceed to
consider the eikonal approximation to the amplitude in a very limited
regime.  We keep only leading-$\lambda$ effects (pure gravity),
$x^\perp\ll 1/\Lambda$ (no effects from confinement), $z\ll z_{max}$
(where the hadrons are small compared to the confinement scale, and
the metric is pure $AdS_5$), but with $z$ and $z'$ large enough that
the proper energy $\sqrt{\tilde s}$ is large compared to the Planck
mass. In addition, the proper distance between the points
$(x^\perp,z)$ and $(x'^\perp, z')$ must not be too small, so that the
scattering involves only linear gravity.  We also ignore the space
$X$, assuming there is no angular-momentum transfer in the compact
directions. 

In the regime where the eikonal approximation is appropriate,
it is easy to adapt flat-space methods to write the eikonal 
result.  This is because $AdS$ spaces have Minkowski slices
with ordinary boost invariance,
and because the derivation of the eikonal
approximation involves separating the light-cone directions from
the transverse directions, which need not be translationally invariant.
(This will be most clear in our perturbative derivation.)
The result for the amputated 5-dimensional amplitude in the
eikonal approximation is
\bea\label{AdSgravityamp}
{\cal A}_{eik}(s,x^\perp,z;y_\perp,w;x'^\perp,z';y_\perp,w)  &=& 
{\cal M}_{eik}(s,x^\perp,z;x'^\perp,z')  \left[g(z)g(z')\right]^{-1/2}
\ \ \ \ \ \ \ 
 \\ &  &  \ \times
\delta^2(x^\perp-y^\perp)\delta(z-w)\delta^2(x'^\perp-y'^\perp)\delta(z'-w')
\nonumber
\eea
showing the classic eikonal reduction of a function of four positions to a
function of two positions in the transverse dimensions, and with
\be
{\cal M}_{eik}(s,x^\perp,z;x'^\perp,z')= -2i\left({z z'\over R^2}\right)^2 s 
\left[\exp\left\{i\chi(s,x^\perp,z;x'^\perp,z')\right\}-1\right]
\ee
where\footnote{Since
$\kappa_5^2/R^3=1/(M_PR)^3$, 
where $M_P$ is the five-dimensional Planck constant, we see that
 $\chi$ depends only on the
number of colors $N$ in the gauge theory, $\chi\sim N^{-2}$, where $N \propto (M_P R)^{3/2}$.}
\bel{AdSgravitychi}
\chi(s,x^\perp,z;x'^\perp,z')= 
 {1\over 2} {\kappa_5^2\over R^3}\ z z' s \ G_3(u)
\ee  
We emphasize again that this form for the amplitude is valid only in the
restricted regime described above.

Our result can now be combined with external wave functions,
normalizable or nonnormalizable, and integrated over $z$ and $z'$, to
allow partial computation of unamputated high-energy scattering
amplitudes, operator matrix elements or Green functions in a dual
gauge theory.  However, since the result above
holds only in limited regions of the coordinates $z, z'$, no complete
physical amplitude can be obtained from this result alone, at least
not without additional arguments showing that all other regions give
small contributions to the full unamputated amplitude.

In the remaining sections we give multiple lines of argument that support
the result (\ref{AdSgravityamp})-(\ref{AdSgravitychi}).

\section{A consistency argument}

First, we check the form of
(\ref{AdSgravityamp})-(\ref{AdSgravitychi}) by matching it to previous
work.  The amputated form for the scattering amplitude in {\it
transverse representation} (longitudinal momentum space and transverse
position space) must agree at transverse proper distances short
compared to $R$ with known results on the eikonal approximation
obtained by other methods, including direct computation of multi-loop
amplitudes at high energy
\cite{CW,Kabat:1992tb,Giudice:2001ce,Amati:1993tb} and the
shock-wave approach \cite{'tHooft:1987rb,'tHooft:1990fr}.

In transverse representation, flat-space results for eikonal
scattering take simple forms.  We know that if the {\it proper}
transverse distance $\tilde b$ between two particles is sufficiently
small compared to $R$, as they scatter at high {\it proper} energy
$\sqrt{\tilde s}$, then standard results must apply.\footnote{Note
that this argument requires that $\tilde s$, though large compared to
the momentum transfer, must not be so large that the eikonal
approximation is nowhere valid for $\tilde b< R$.  For fixed $\tilde
s$ we can always consider taking $R$ sufficiently large that the
argument applies; as we will see this is enough to fix the
answer. Similarly $\tilde b$ must not be so small as to probe the
nonlinear gravity near the scattering objects.}  In $D= 5$ (bulk)
spacetime dimensions, the leading order amputated amplitude in
transverse representation (with transverse delta functions removed)
will be
\bel{flatleadingorder} 
{\cal M}^{(1)}(\tilde s,\tilde b) 
= {1\over\sqrt{g_{+-}g_{-+}}} \kappa_5^2
{\tilde s^2\over 4 \pi \tilde b} 
\ee
The functional dependence on $\tilde s$ and $\tilde b$ is 
as in flat space; the
metric factor in front is due\footnote{More precisely, though natural,
it is actually
conventional, since one could absorb it into the amputation
prescription.} to
our use of a transverse position basis.  The flat-space
eikonal approximation then implies
\bel{flatlimitTfull}
{\cal M}_{eik}(\tilde s,\tilde b) 
\propto -2i{\tilde s \over\sqrt{g_{+-}g_{-+}}} 
\left[\exp\left\{{i\over 2}\kappa_5^2
{\tilde s\over 4 \pi  \tilde b}\right\} - 1 \right]\ .
\ee
Here we have used the fact that every order of the eikonal approximation
must transform in the same way under diffeomorphisms, so the
phase shift in the exponent must be an invariant.  

On the other hand, we also know that for $\lambda \to\infty$, the
leading-order high-energy scattering amplitude in $AdS_5$ is due
simply to $t$-channel graviton exchange, so the gravitational
propagator ${\cal G}_{MN,M'N'}$ must appear in the tree amplitude.  At
high energy, only the term in ${\cal G}$ proportional to $g_{++}
g_{--}$ survives, and  this term is just the massless scalar
propagator $G_5$ in $AdS_5$, a function by $AdS$ isometries
(the conformal invariance of the dual gauge theory) of the chordal
distance
\be
u_5 = {\eta_{\mu\nu}(x-x')^\mu(x-x')^\nu + (z-z')^2\over2 z z'}
\ee
Explicitly \cite{Freedman,D'Hoker:1999jc}
\be
{\cal G}_{++,--}(x,z;x',z') = 
 2 \left({R^2\over z z'}\right)^2 G_5(u_5) .
\ee
with 
\be
G_5(u_5)= {1\over 8\pi^2}\left[2+{1+u_5\over[u_5(2+u_5)]^{3/2}} 
- {2(1+u_5)\over[u_5(2+u_5)]^{1/2}}\right]
\ee

For near-forward scattering at high energy, the total momentum transfer
is limited. In a collinear frame where the light-cone 
momentum components become large, 
the total momentum transfer is nearly transverse and the
longitudinal components of momentum transfer must go to zero
asymptotically. 
In such a frame,  the $AdS_5$ 
massless scalar propagator $G_5$ should be evaluated
at zero longitudinal momentum transfer, {i.e.}
\bel{intG0}
G_5(q_{\pm}=0, x^\perp,z;x'^\perp,z') = 
\int dx^+ dx^- G_5(u_5) = z z' G_3(u)
\ee
since $u_5=u -   x^+x^-/z z'$.  
Here we used the definitions in (\ref{Gfreedman})
and (\ref{udef}).\footnote{Note this is a special case of a general 
relation, which states that the
propagator of a bulk scalar in $D$ dimensions of mass $(mR)^2 =
\Delta(\Delta-D+1)$, at zero longitudinal momentum transfer, is $zz'$
times the propagator for a bulk scalar in $AdS_{D-2}$ of mass $(mR)^2
= (\Delta-1)(\Delta-D+2)$.  This is dual to the following simple
statement.  The two-point function for an operator of dimension
$\Delta$ in $D-1$ dimensions on the boundary of $AdS_D$, at zero
longitudinal momentum transfer, reduces  to the
two-point function of an operator of dimension $\Delta-1$ in $D-3$
boundary dimensions.}

Now we may simply note that a scalar propagator in 5 dimensions
approaches $1/\tilde b$ as $\tilde b\to 0$, since the propagator
satisfies a flat transverse Laplacian at short distances.  Observing
that $\tilde b\to R\sqrt{2u}$ as $u\to 0$ and matching to
(\ref{flatleadingorder}) and (\ref{flatlimitTfull}) fixes the results
\bel{leadingorder} 
{\cal M}^{(1)}(\tilde s,x^\perp,z;x'^\perp,z') = 
{1\over R} \left(zz's\over R^2\right)^{2} \left(zz'\over R^2\right)
G_3(x^\perp,z;x'^\perp,z')
\ee
and
\bea\label{T5d}
{\cal M}_{eik}  & = &  
-2i {1\over \sqrt{g_{+-}g_{-+}}} \tilde s 
\left[\exp\left\{{i\over 2} \kappa_5^2 R^{-1}
{\tilde s \ G_3(u)}\right\} -1\right]\ \nonumber \\
& = & \
-2is \left({z z'\over R^2} \right)^2 \left[\exp\left\{{i\over 2} 
{\kappa_5^2\over R^{3}}
{z z' s \ G_3(u)}\right\} -1\right]\ ,
\eea
in agreement with our earlier claim.

The above arguments can all be easily generalized to other dimensions.
The only work is to obtain the correct normalization.

\section{A Diagrammatic Derivation}
\label{sec:perturbative}

\begin{figure}[h]
\begin{center}
\includegraphics[width = 0.8\textwidth]{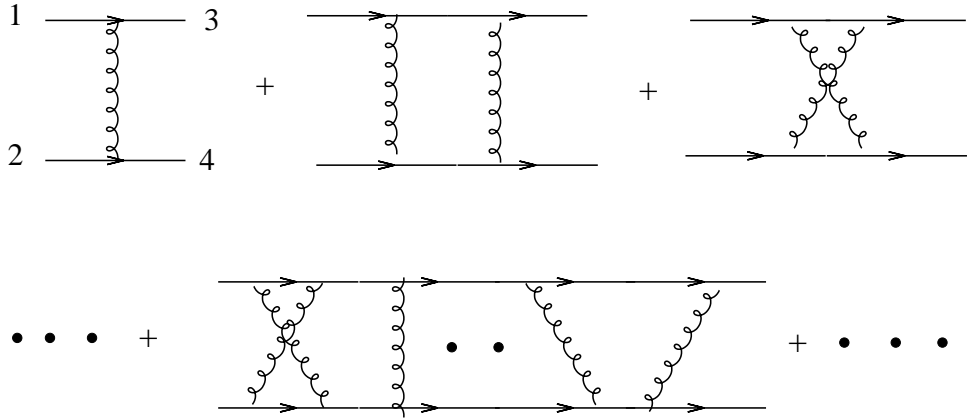}
\caption{Ladder and crossed ladder diagrams contributing to the eikonal approximation in the high    energy limit.}
\label{fig:eik}
\end{center}
\end{figure}

We may also derive the eikonal result as a sum of
the high-energy contribution of perturbative diagrams illustrated in
Fig.~\ref{fig:eik}, along the lines of 
Cheng and Wu \cite{CW,Chang:1971je,Levy:1969,Amati:1987wq,Fried:1990,Fried:2002ds} or the
methods pioneered by many other authors in the gravitational case; see
for example \cite{'tHooft:1987rb,'tHooft:1990fr,Kabat:1992tb,Giudice:2001ce,Amati:1993tb,Amati:1992zb,Kabat:1992pz}.  For Anti de Sitter space this consists of summing
a class of Witten diagrams, where we choose scalar fields for the
external lines and gravitons for the exchanged rungs between these two
sides.  The sum includes all orders of the coupling to the sources
giving rise to all ladder and crossed ladder diagrams. The mechanism
leading to eikonalization at high energies in flat background has long
been understood. From the perspective of perturbative summation, the
key simplification necessary is the separation of dependence on the
{\it longitudinal} light-cone momentum coordinates, $q^{\pm}_i$ and
the {\it transverse} impact $x^\perp_i$ co-ordinates.  This feature
can already be appreciated by analyzing the high-energy behavior for
the sum of box and crossed-box diagrams in flat space.  Therefore, we
begin by providing a brief description in $\phi^3$
theory,
paying particular attention to the dependence of amplitudes on transverse
coordinates,  before treating the case of graviton exchange in $AdS_5$.

\subsection{Flat Background}

We begin by considering the 4-particle amputated Green's function,
${\cal A}(p_i)$, for elastic scattering $p_1 + p_2 \rightarrow (-p_3) +
(-p_4)$, by exchange of a massless field of spin $J$. We use the
``all-incoming" convention, with $s=-(p_1+p_2)^2$, $t=-q^2=- (p_1+p_3)^2$.
The reasons for starting with the Green's function rather than an
on-shell scattering amplitude are twofold. First, in the conformal case
there is no on-shell S matrix, and second, we choose to transform all
transverse momenta $p^\perp_i$ to co-ordinate space, which is not
possible on the mass shell ($p^2_i = - m^2$).  This procedure greatly
simplifies the analysis of the high-energy dependence of the ladder
graphs.  After this transformation to  the ``transverse
representation'', we also choose to drop the overall conservation
delta function for longitudinal momentum: $(2 \pi)^2 \delta(p^\pm_1 +
p^\pm_2 + p^\pm_3 +p^\pm_4)$.  In the high-energy limit where
$p_1^+\simeq -p^+_3$ and $p_2^-\simeq -p^-_4$ are large, we will
verify that, order by order, the scattering is local in
transverse coordinates,
\be
{\cal A}(p_i^\pm, x_i^\perp)=\sum_{n=1} {\cal A}^{(n)} \rightarrow \left(\sum_n{\cal M}^{(n)} (p^+_1, p^-_2, x_1^\perp - x_2^\perp)\right)\delta^2(x^\perp_1-x^\perp_3)\; \delta^2(x^\perp_2-x^\perp_4)
\ee
and that the resulting sum, 
\be
{\cal M}_{eik} (p^+_1, p^-_2, x^\perp - x'^\perp)=\sum_{n=1} {\cal M}^{(n)} (p^+_1, p^-_2, x^\perp -   x'^\perp) = -2is \; \left[e^{i \chi(s,x^\perp - x'^\perp)} -1\right] \; ,
\ee
takes the eikonal form.

\subsubsection{Tree and One-loop Scattering in a Flat Background}

The tree level amputated amplitude in transverse representation, 
\be
{\cal A}^{(1)} ( p_i^\pm, x^\perp_i) = {\cal M}^{(1)} ( p_i^\pm, x^\perp_1 - x^\perp_2) \delta^{2}(x_1^\perp-x_3^\perp) \delta^{2}(x^\perp_2-x_4^\perp) 
\ee
\be
{\cal M}^{(1)} ( p_i^\pm, x^\perp_1-x^\perp_2) =g_0^{2 } s^J G(q^\pm, x^\perp_1-x^\perp_2) 
\ee
is given in terms of the $t$-channel massless propagator,
\be
 G(q^\pm, x^\perp)=  \int \frac{d^2q^\perp}{(2\pi)^{2}} \frac { e^{i q^\perp x^\perp }      }{ {q^\perp}^2  -2 q^+q^-   -i\epsilon   } \; ,
\ee
 where $q^\pm= - (p_1^\pm+p_3^\pm)$. We have  introduced a factor
$s^J$ for each $t$-channel exchange in anticipation of the
case of a graviton exchange, where $J=2$; of course our scalar model has
$J=0$.  At high energies, $q^{\pm}=O(1/\sqrt s)$, we have
\be
{\cal M}^{(1)} ( p_1^+,p_2^-, x^\perp-x'^\perp) \simeq  g_0^{2 } s^J G(q^\pm=0, x^\perp-x'^\perp)  \; .
\ee
This is to be compared with Eq.~(\ref{eq:gravitontree}) for the 
one-graviton exchange contribution in $AdS_5$.

\begin{figure}
\begin{center}
\includegraphics[width = 0.9\textwidth]{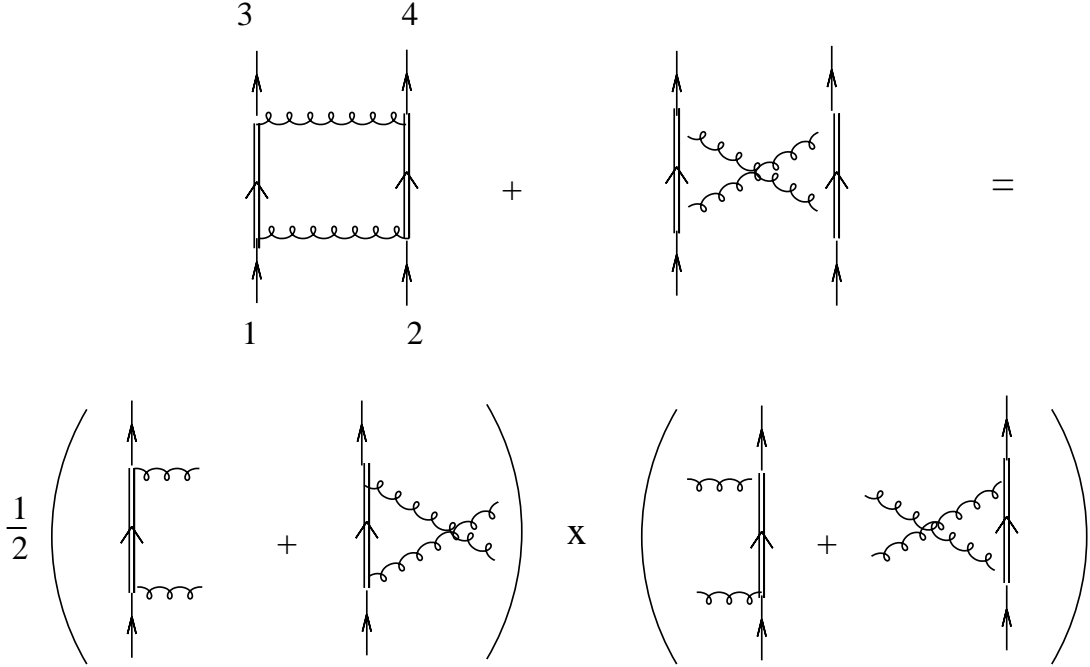}
\caption{Sum of box and cross box diagram is factorized with combinatoric
weight $1/2!$.}
\label{fig:box}
\end{center}
\end{figure}

The amputated amplitude at one-loop order involves a box diagram, and
a crossed box diagram, obtained from the box 
by interchanging $(p^\pm_1,x_1^\perp)$
with $(p^\pm_3,x_3^\perp)$ or equivalently $(p^\pm_2,x_2^\perp)$ and
$(p^\pm_4,x_4^\perp)$.  The sum of the two diagrams, in transverse
representation, can be written in a compact form (see
Fig~\ref{fig:box}):
\bea
&& {\cal A}^{(2)}(p_i^\pm, x_1^\perp,x_3^\perp,x_2^\perp,x_4^\perp)=  \frac{i (2\pi)^2}{ 2!\;  } \int 
\frac{d^2q^\pm_1}{(2\pi)^2}  \frac{d^2q^\pm_2}{(2\pi)^2} \delta^2(q^\pm-q^\pm_1-q^\pm_2) A_{13}(p_1^\pm, q_1^\pm, x_1^\perp, x_3^\perp)  \nonumber\\
&&\times  [(-i g_0^2 s^{J}) G(q_1^\pm, x_1^\perp-x_2^\perp)] [(-i g_0^2 s^J) G(q_2^\pm, x_3^\perp-x_4^\perp)] A_{24}(p_2^\pm,- q_1^\pm, x_2^\perp, x_4^\perp)
\label{eq:scalarflatboxkernel}
\eea
\bea
A_{13}(p_1^\pm, q_1^\pm, x_1^\perp, x_3^\perp) &=&  S(p^\pm_1+q^\pm_1, x_3^\perp,x_1^\perp) + S(p^\pm_1+q^\pm_2, x_1^\perp,x_3^\perp)\nonumber\\
 A_{24}(p_2^\pm, q_1^\pm, x_2^\perp, x_4^\perp) &=&  S(p^\pm_2+q^\pm_1, x_4^\perp,x_2^\perp) + S(p^\pm_2+q^\pm_2, x_2^\perp,x_4^\perp)  \label{eq:Boxpermutation}
\eea
Here  $S$ is the propagator for the scattered particles, and for a scalar of mass $\mu$, it is 
\be
 S(p^\pm, x^\perp)=  \int \frac{d^2p^\perp}{(2\pi)^{2}} \frac { e^{i p^\perp x^\perp}      }{ {p^\perp}^2  -2 p^+p^-      +\mu^2 -i\epsilon } \; .
\ee
The two terms on the right-hand side
of each equation in (\ref{eq:Boxpermutation}) represent
a sum over all possible orderings of 
the rungs of the ladder as they connect
to one of the ladder's sides. 
The product of $ A_{13}$ and $ A_{24}$ in Eq.
(\ref{eq:scalarflatboxkernel}) then leads to four terms.  Since the rungs
are indistinguishable, these four terms represent a double-counting
of each Feynman diagram. Consequently 
a factor of $1/2!$ has been added to compensate for this
over-counting.

Extracting the high-energy behavior of
Eq.~(\ref{eq:scalarflatboxkernel}) can be done in the following two
steps.  We first note that, for near-forward scattering at high
energies, the limit $s$ large is characterized by $p_1^+\simeq -p_3^+$
and ${p_2}^-\simeq -p_4^-$ both large, and $q$, $q_1$, $q_2$
asymptotically space-like, with $q_{i}^{\pm} =O(1/\sqrt s)$. In this
limit, $ G(q_1^\pm, x_1^\perp-x_2^\perp)\simeq G(q_1^\pm = 0,
x_1^\perp- x_2^\perp)$ and $ G(q_2^\pm, x_3^\perp-x_4^\perp)\simeq G(q_2^\pm=0,
x_3^\perp-x_4^\perp)$ can be taken out of the $\int d^2q^\pm_1
d^2q_2^\pm \delta^2(q^\pm-q^\pm_1-q^\pm_2)$ integrals.

For definiteness, let us next use
 $q_{1}^\pm$  as independent
integration variables. At high energies, the dependence of $ A_{13}$ on
$q_1^+$ drops out and it becomes a  function of $q_1^-$ only.
Conversely, $A_{24}$ is independent of $q_1^-$ and is a 
function of $q_1^+$. This factorizable dependence, $ \int  d^2q^-_1  \int d q_1^+   A_{13}  A_{24} \simeq \int  d^2q^-_1   A_{13}\int d q_1^+ A_{24}$,       allows
us to carry out the $q_1^{\pm}$ integrations explicitly.
Focus first on the integral over $ A_{13}$, 
\be
 \int  \frac{d q^-_1}{2\pi}  A_{13}(p_1^+, q_1^-, x_1^\perp, x_3^\perp)   
\ee
which involves a sum of two pole terms coming from the $S$ propagators. At
high energies, the $s$-channel pole occurs at $0=\mu^2+
(p_1+q_1)^2-i\epsilon$, i.e., $q_1^- \simeq O(1/p_1^+) -i \epsilon $, and the u-channel pole occurs at $q_1^- \simeq O(1/p_1^+) + i
\epsilon$.  Although
each pole term vanishes only as $O(1/q_1^-)$ for $q_1^-$ large, the sum
goes as $O(1/{(q_1^-)}^2).$ Closing the contour leads simply to 
\be
(-i)\;{\rm Residue} \; [\;S\;]_{q^-\simeq 0-i\epsilon} = (i /2 p_1^+) \int \frac{d^2p^\perp  e^{i p^\perp (x_1^\perp- {x_3}^\perp) }   }{(2\pi)^2}     =(i  /2 p_1^+ ) \delta^2 (x_1^\perp- {x_3}^\perp)
\ee
Similarly for 
$ A_{24}$,  closing the contour in $q_1^+$ leads to
$(i /2 p_2^- )\delta^{2}(x_2^\perp- {x_4}^\perp)$.
Putting these together, we wind up with
\be
{\cal A}^{(2)}(p_1^+,p_2^-, x_1^\perp,x_2^\perp,x_3^\perp,x_4^\perp)
\simeq{\cal M}^{(2)}(p_1^+,p_2^-, x_1^\perp-x_2^\perp) \delta^2(x_1^\perp- x_3^\perp)  \delta^2(x_2^\perp- x_4^\perp)
\ee
\be
{\cal M}^{(2)}(p_1^+,p_2^-, x^\perp-x'^\perp) = -2i s \; \frac{1}{2!}\;  
[i g_0^2 s^{J-1}  G(q^\pm=0, x^\perp-x'^\perp)/2]^2\;. \label{eq:boxHEflat}
\ee
Note that all the dependences on $x_1^\perp-x_3^\perp$ and
$x_2^\perp-x_4^\perp$ reduce to delta-functions, i.e., the effective
interaction remains local, with ``zero transverse deflection". This
is a key feature common to all eikonal results, and
we will see in a moment 
that it generalizes to the case of $AdS$ space.\footnote{It is worth providing a more intuitive interpretation of the result just
obtained. The $q_1^-$ integral over $ A_{13}$ can be written more
symmetrically as
$$
\int dq_1^-dq_2^-\delta(q_1^- + q_2^-)  A_{13}
\sim
\frac{1}{-2p_1^+}\int dq_1^-dq_2^-\delta(q_1^-+q_2^-)
\times
\left[\frac {1}{q_1^-+i\epsilon} + \frac {1}{q_2^-+ i\epsilon}\right]
\label{eq:timeordering}
$$
The integrand can be shown simply to correspond to the Fourier transform
of $\theta(x_3^+-x_1^+) + \theta(x_1^+-x_3^+) $. That is, the
different permutations in Eq.~(\ref{eq:Boxpermutation}) simply correspond 
to scattering in different
``time-orderings".  For each ordering, the scattering amplitude is constant and
local, proportional to $(1/p_1^+)\delta^2(x_1^\perp-x_3^\perp)$. This
physical picture generalizes to the case of multiple exchanges.}

\subsubsection{Eikonal Exponentiation }

We can now re-sum the infinite series of loop graphs to obtain the
eikonal approximation, as in \cite{CW}.  Consider a ladder with $n$
rungs, where arbitrary crossings of rungs are allowed. Denote
transverse coordinates for these $2n$ vertices by $\{x_i\}$ and
$\{x'_j\}$, $i,j=1,\cdots, n$. The sum of all $n^{th}$ order diagrams
can be obtained by evaluating
\bea
&& \frac{ i (2\pi)^2}{   n!  }  \int  \prod_{k=1}^n\frac{d q^+_k dq_k^-}{(2\pi)^2} \delta^2\left(\sum q^\pm_i\right)A_{13} \left(p^\pm_1,p^\pm_3, q^\pm_i, x_j^\perp  \right)  \nonumber\\
&\times&     \left[\prod_k  \left( -i  g_0^2 s^J\right) G\left(q^\pm_k\simeq 0 , x^\perp_k, x'^\perp_k\right)\right]   A_{24} \left(p^\pm_2, p^\pm_4, -q^\pm_l,  x'^\perp_p \right)
\eea
where we have ordered the light-cone 
momenta of exchanged rungs, $q^\pm_k$, $k=1,\cdots, n$,  and 
have also dropped the  dependence of
exchanged propagators on $\{q^\pm_k\}$, as is valid at high energies, 
and as was done earlier for the one-loop contribution.
Here $A_{13}$ is given by a sum of $n!$ terms, each 
a product of $n-1$ propagators, corresponding 
to all possible  ways of attaching $n$
exchanged propagators to one side of the eikonal ladder.
$ A_{24}$ is defined similarly.  Both are
generalization of Eq.~(\ref{eq:Boxpermutation}) 
 from $n=2$ to $n>2$. As
for the $n=2$ case, a
factor of $1/n!$ is supplied to account for over-counting.

To extract the high-energy behavior, we again take advantage of
$\{q^+_i\}$ and $\{q^-_i \}$ factorization, which allows us to carry out the
integrations
\be
  \prod_i \int \frac {dq^-_i}{2\pi}   \delta(\sum_iq_i^-)  A_{13} (p^+_1, q^-_i, x_j^\perp)         \int \frac{dq^+_i}{2\pi} \delta(\sum_iq_i^{+})   A_{24} (p^-_2, -q^+_k, x'^\perp_l)
\ee
Following the analysis of Cheng and Wu \cite{CW}, one finds that the net result of these integrations  is to produce 
\be
 [ (i)^2/2s]^{n-1} \prod_{i=1}^{n-1}[\delta^{2}(x^\perp_i-x^\perp_{i+1})\delta^{2}(x'^\perp_i-x'^\perp_{i+1})] \;.
\ee
That is, we can set $x^\perp_i=x^\perp$ and $x'^\perp_j=x'^\perp$, and the feature of ``zero transverse deflection" persists in each order. Integrating over all but 2 of these transverse coordinates, we are led to
\be
{\cal A}^{(n)}(p_1^+,p_2^-, x_1^\perp,x_3^\perp,x_2^\perp,x_4^\perp)\simeq {\cal M}^{(n)}  ( p^+_1,p^-_2,x_1^\perp,x_2^\perp ) \delta^{2}(x_1^\perp- x_3^\perp)  \delta^{2}(x_2^\perp- x_4^\perp)\label{eq:scalarflatnth}
\ee
\be
{\cal M}^{(n)} ( p^+_1,p^-_2,x^\perp-x'^\perp )  = -  2is \; \frac{1}{n!}\; [ i g_0^2 s^{J-1}   G(q^\pm=0, x^\perp-x'^\perp)/2 ]^n
\ee

After summing over $n$,  we arrive at the eikonal amplitude
\be
{\cal M}_{eik}( p^+_1,p^-_2,x^\perp-x'^\perp )=\sum_{n=1} {\cal M}^{(n)}( p^+_1,p^-_2,x^\perp-x'^\perp )=  ( - 2 is  )  [ e^{ i \chi(s,x^\perp - x'^\perp)}-1  ] 
\ee
where the eikonal is
\be
\chi( s,x^\perp- x'^\perp)= \frac{1}{2}\; { g_0^2} \ s^{J-1}\  G(q^\pm=0, x^\perp-x'^\perp)
\ee
Upon taking a 2-d Fourier transform, we arrive at the on-shell amplitude
\be
T(s,t) =  \int d^2x^\perp e^{- ix^\perp\cdot q^\perp}{\cal M}( p^+_1,p^-_2,x^\perp) \simeq     - 2 is    \int d^2x^\perp e^{- ix^\perp\cdot q^\perp} [ e^{ i \chi(s,x^\perp - x'^\perp)}-1  ] 
\ee

\subsection{Eikonal Expansion for $AdS_5$  gravity }

Let us return to the problem of summing eikonal graphs in $AdS_5$,
which can be carried out in close analogy with the flat background.
As described earlier, we begin by considering a gauge theory
scattering amplitude (or correlation function), truncated by dropping
the external hadron wave functions (or external boundary-to-bulk
$AdS_5$ propagators) on each external leg, and then written in the
transverse representation $(p^\pm,x_i^\perp,z_i)$. We work only in the
regime where the amputated amplitude can be evaluated using
propagators in $AdS_5$ space, which are conformal Green's function in
$AdS_5$ with 3 transverse dimensions in an $AdS_3$ submanifold.

In the high-energy limit we only need to keep the $++,--$ component of
the graviton propagator, which simplifies the analysis greatly. In
what follows, we generalize this to $n$-graviton exchanges and
observe how eikonalization arises for scattering in $AdS_5$.

Consider  the case of the one-graviton-exchange 
Witten diagram for scalar sources on the boundary of $AdS_5$. 
The amplitude for this diagram is \cite{Freedman}
\be
 {\kappa_5^2}\int  dz \sqrt g \int d z' \sqrt {g'} \; \tilde T^{MN}(p_1,p_3, z) \tilde G_{MNM'N'}(q,z,z') \tilde T^{M'N'}(p_2, p_4, z')  \label{eq:qonegraviton}
\ee
in momentum representation,
where $\tilde T^{MN}$ is the energy-momentum tensor for the scalar
source in the bulk and $\tilde G_{MNM'N'}$ is the graviton propagator,
both in momentum representation.  At high energies, keeping the
leading $++,--$ component, we find for the amputated amplitude in
transverse representation
\be
{\cal M}^{(1)}(s, x^\perp,z;x'^\perp,z') =  \frac{\kappa_5^2}{R} \;  { \tilde s}^2\; \left( \frac{zz'}{R^2}\right)   \;G_3(x^\perp, z, x'^\perp, z')  \label{eq:onegraviton}
\ee
which has previously been given in Eq.~(\ref{eq:gravitontree}). Here $R$ is the $AdS$ radius and $G_3$ is the $AdS_3$ scalar propagator, Eq.~(\ref{Gfreedman}), which can be expressed in terms of the chordal distance, Eq.~(\ref{udef}).

Let us turn next to the one-loop contribution, which involves a box diagram and a crossed box.  The total contribution at high energies, generalizing
Eq.~(\ref{eq:scalarflatboxkernel}) 
by keeping only graviton
exchanges of helicity structure $(++,--)$, can be expressed as
\bea
{\cal  A}^{(2)} ( p^\pm_i,x_i^\perp,z_i)&=&  i \frac{ 1 }{2! (2\pi)^2 }  \int 
d^2q^\pm_1  d^2q_2^\pm \delta^2(q^\pm-q^\pm_1-q^\pm_2) \nonumber\\
&\times& [A_{13}] \;\left[ -i  (\kappa^2_5/R^3)(z_1z_2s/R^2)^2(z_1 z_2) 
G_3(u[1,2])\right] \nonumber\\
&\times &  
\left[-i (\kappa^2_5/R^3)(z_3z_4 s/R^2 )^2 (z_3z_4) G_3(u[3,4])\right]\; [A_{24}] 
\eea
where $u[1,2]$ and $u[3,4]$ are chordal distances in an obvious notation. Again, similar to Eq.~(\ref{eq:Boxpermutation}), we have
\bea
&& A_{13} (p^\pm_1, q^\pm_1,q_2^\pm,x_1^\perp, z_1,x_3^\perp, z_3) \nonumber\\
 &&\quad\quad\quad\quad\quad\quad= {1\over R^3}\left[G_5(p^\pm_1 + q^\pm_1, x_3^\perp,z_3,x_1^\perp,z_1) + G_5(p^\pm_1 + q^\pm_2, x_1^\perp, z_1,x_3^\perp,z_3)\right] \nonumber\\
&& A_{24} (p^\pm_2, -q^\pm_1,-q_2^\pm,x_2^\perp, z_2,x_4^\perp, z_4) \nonumber\\
&&\quad\quad\quad\quad\quad\quad= {1\over R^3}\left[G_5(p^\pm_1 - q^\pm_1, x_4^\perp,z_4,x_2^\perp,z_2) +  G_5(p^\pm_1 - q^\pm_2, x_2^\perp, z_2,x_4^\perp,z_4)\right ]  \nonumber\\
  \label{eq:permutations}
\eea
which account for both the ``box"  and  the ``cross-box" diagrams. Here $G_5(p^\pm, x^\perp,z;x'^\perp,z')$ is the $AdS_5$ scalar propagator in a transverse representation, which has previously been introduced, Eq.~(\ref{intG0}).
It can be expressed as
\be
G_5(p^\pm, x^\perp,z;x'^\perp,z')= (zz')^2\int \frac {d^2 p^\perp }{ (2\pi)^2}e^{ip^\perp (x^\perp-x'^\perp)} \int k dk \frac {J_2(zk) J_2(z'k)}{ k^2 +{ p^\perp}^2 -2 p^+p^-} \label{eq:G5mixedrep} \ .
\ee

Let us concentrate on extracting the high-energy behavior for ${\cal
  A}^{(2)} $. The situation is nearly identical to that for a flat
background, leading to factorization in $q^\pm_1$, and the need to
evaluate $(1/2\pi)\int dq_1^- A_{13}$ and $(1/2\pi)\int d q^+ A_{24}$
separately. Focus on the $A_{13}$ integral, which again involves
two terms, each an integral over a propagator $G_5$. Using 
Eq.~(\ref{eq:G5mixedrep}), integration over $q_1^-$ leads to
\bea
 {- i \over R^{3} }{\rm Residue} \; [\;G_5\;]_{q^-\simeq 0-i\epsilon} &= &(iR^{-3}  /2 p_1^+) \int \frac{d^2p^\perp}{(2\pi)^2}  e^{i p^\perp (x_1^\perp- x_3^\perp) }    (z_1z_3)^2\int k dk  J_2(z_1k)J_2(z_3k)   \nonumber\\
 & =& (i R^2 /2 p_1^+z_1^2) \delta^2 (x_1^\perp- x_3^\perp)\delta(z_1-z_3)/\sqrt g_1\;
\eea
using the Bessel function completeness relation; here $g_1\equiv {\rm det} \; g(z_1)$.
Similarly, we obtain
\be
(1/2\pi)\int  dq_1^+  A_{24} = (iR^2/2 p_2^-z_2^2)    \delta^2 (x_2^\perp- x_4^\perp)\delta(z_2-z_4)/\sqrt g_2\; .
\ee
Putting these together, we again verify ``zero transverse deflection", and
\be
{\cal M}^{(2)} ( p_1^+,p_2^-,x^\perp,z;x'^\perp, z')   = - 2 i (zz'/R^2)^2s \; \frac{1}{ 2!} \;  [  i (\kappa^2_5/2R^3 ) (zz's) G_3(u)]^2
 \label{eq:eikonalAdSamplitude2}
\ee
This represents a direct generalization of the flat-space result, Eq.~(\ref{eq:boxHEflat}).

The generalization to higher loops can similarly carried out as for flat space. We obtain
\bea
{\cal A}^{(n)}(p_i^\pm, x_i^\perp,z_i)&\simeq&  {\cal M}^{(n)}(s,x_1^\perp,z_1,x_2^\perp, z_2)\nonumber\\
&\times & [ \delta^2(x^\perp_1-x^\perp_3)\delta(z_1-z_3)/\sqrt g_1][\delta^2(x^\perp_2-x^\perp_4) \delta(z_2-z_4)/\sqrt g_2]
\eea
\be
{\cal M}^{(n)} ( p_1^+,p_2^-,x^\perp,z;x'^\perp, z')   = - 2 i (zz'/R^2)^2s \; \frac{1}{ n!} \; \left[  i (\kappa^2_5/2R^3) (zz's) G_3(u)\right]^n
 \label{eq:eikonalAdSamplituden}
\ee
Summing over $n$, we have
\be
{\cal M}_{eik}( p_1^+,p_2^-,x^\perp,z;x'^\perp, z')=- 2 i s\; \left(\frac{zz'}{R^2}\right)^2 \left[ e^{ i \chi(s,x^\perp-x'^\perp ,z,z')}-1\right]
\ee
where
\be
\chi(s,x^\perp-x'^\perp ,z,z')= \frac{1}{2}\; \left(\frac{\kappa^2_5}{R^3}\right) (zz's) \;G_3(u)=  \frac{1}{2}\; \; (R M_{P})^{-3}(zz's) \;G_3(u)
\ee
as promised.

Lets up summarize the essential feature of the eikonal approximation.
The dependences on $x_1^\perp-x_3^\perp$, $z_1-z_3$ and
$x_2^\perp-x_4^\perp$, $z_1-z_3$ reduce to delta-functions
so that there is ``zero transverse deflection" of the incoming states
during the interaction. As we now will see in the
shock wave derivation, this ``freezing'' of transverse motion is a
consequence of the instantaneous interaction in light-cone time $x^+$.

\section{Shock Wave Derivation}

An alternative approach to the eikonal approximation for gravity is to
study the semi-classical limit of one particle scattering in the
presence of a shock wave created by the other. In particular 't Hooft
computed the eikonal amplitude for high-energy scalar particles in
flat space gravity~\cite{'tHooft:1987rb,'tHooft:1990fr}. The shock is
given by the Aichelburg-Sexl metric~\cite{Aichelburg:1970dh}, which is
the Schwarzschild metric for a particle with mass $m_i \ll M_{P}$ boosted to
the light-cone and approximated for impact parameters far outside the
Schwarzschild radius. 
Here we will show that by generalizing this
metric to a shock wave in the bulk 5-d $AdS$ space~\cite{Arcioni:2001my,Kang:2004jd,Kang:2005bj}, we are able to
derive the eikonal amplitude without recourse to perturbation theory
used in Sec.~\ref{sec:perturbative}. This has the advantage that it
provides greater
insight and a complementary way to understand the source of
corrections to this approximation.

Consider the shock wave created by particle 2 with a very large
longitudinal light-cone ``energy'' $p^-_2$, on the 
light-cone $x^+ = (x^0 + x^3)/\sqrt{2} = 0$ at fixed transverse
position $x^\perp = x'^\perp$ and $z  = z'$. 
The  energy momentum tensor for this particle in the {\it bulk} is  
\be
 T^{--}(x^+,x^\perp,z;x'^\perp,z') = (z^2/R^2) p^-_2 \delta(x^+) \delta^2(x^\perp- x'^\perp) \delta(z - z')/\sqrt{g} \; .
\label{eq:source}
\ee
Although tensor indices are raised and lowered by the $AdS$ background
metric $g_{MN} = \eta_{MN}R^2/z^2$, we choose to treat the momentum
components, $p^\mu = \eta^{\mu \nu} p_\nu$, as flat space 4-vectors,
to match with the Noether currents on the boundary Yang-Mills theory.
Note that extra factors of $z$ in $T^{--}$ ensure that $T_{++} =
g_{+-} g_{+-} T^{--}$ and $h_{++}$ both scale like $z^{-2}$ under
$z\to\gamma z, x^\mu\to\gamma x^\mu, p_\mu\to\gamma^{-1}p_\mu$, as
they must for a conformal dual gauge theory.  With this as the source
to the Einstein equation, one arrives at the modified metric,
\be
ds^2 = (g_{MN} + h_{NM})dx^M dx^N = 
R^2 \; \frac{- 2 dx^+ dx^- + (dx^\perp)^2 + dz^2}{z^2} +  h_{++}(x^+,x^\perp, z)  dx^+ dx^+ \; ,
\ee
where $\sqrt{g} = R^5/z^5$ and $g_{+-} = g_{-+} = - R^2/z^2$. 

The eikonal approximation requires solving gravity in the Gaussian
approximation for fluctuations $h_{MN}$ relative to the fixed $AdS_5$
background metric, $g_{MN}(z)$.  Expanding the Einstein Hilbert action to
quadratic order for the relevant terms we have,  
\bea S_{EH}[g+h] \simeq S_{EH}[g] 
&+& \frac{1}{2 \kappa^2_5} \int
d^4x dz \sqrt{g} \; \left[ - \frac{1}{2}\;\dd_N {h^+}_-\ \dd^N {h_+}^- - \frac{1}{2}\;\dd_N {h^-}_+\ \dd^N {h_-}^+ \right] \nn 
&+&    \int d^4x dz \sqrt{g} \; \left[
 h_{++} T^{++}  +   h_{--} T^
{--} \right]  +  O(h^3) \; , 
\eea
where for convenience we have introduced the dependent metric
functions: ${h^-}_+ = {h_+}^- = g^{-+}h_{++}$ and ${h^+}_- = {h_-}^+ =
g^{+-}h_{--}$. This leads to the linearized Einstein equation\footnote{Note 
that in flat space, we would need to solve for the
  transverse Greens function, $$- \nabla^2_\perp h_{++} =2 \kappa^2_D
  p^-_2\delta(x^+) \delta^{D-2}(x^\perp)\; , $$
  which for $D=4$ agrees
  with the Aichelburg-Sexl metric: $h_{++} = - p^-_2 \kappa^2_4
  \log(|x^\perp|/C) \delta(x^+)/\pi$.},
\be
-\Delta_2\; h_{++}(x^+,x^\perp, z) =  2 \kappa^2_5 T_{++}(x^+,x^\perp, z;x'^\perp,z') \; ,
\label{eq:hpp}
\ee
where $\kappa^2_5 = 1/M^3_P$ and  
\be
\Delta_j=z^{-j} \frac{1}{\sqrt g} \partial_M\sqrt g g^{MN} \partial_N z^j = 
\frac{1}{R^2} [ z^2 \dd_z^2 + (2j -3) z\dd_z  +j (j-4)  + z^2 \nabla^2_\perp] \; ,
\ee
is the general tensor Laplacian operator for $AdS_5$ defined in
Ref.~\cite{BPST}.
The solution to the Einstein equation (\ref{eq:hpp}) is proportional to
the bulk-to-bulk scalar propagator in $AdS_3$:
\be
 h_{++}( x^+,x^\perp,z;x'^\perp,z') =  2 (z'/z) (\kappa^2_5/R) p^-_2 G_3(x^\perp-x'^\perp,z,z')  \delta(x^+) \; ,
\ee
where we have reinserted the explicit dependence on the location,
$(x'^\perp,z')$, of the source (\ref{eq:source}).  Note that the factor of
$z'/z$ is uniquely determined at this point resulting below
in a scattering phase symmetric in $z,z'$. From this solution we
also obtain by the raising operator $g^{-+} = - z^2/R^2$,
\be
 h^{--}( x^+,x^\perp,z;x'^\perp,z') =  2 z z' (\kappa^2_5/R^3)  (z^2/R^2) p^-_2 G_3(x^\perp-x'^\perp,z,z')  \delta(x^+) \; .
\ee
Since the $AdS_3$ propagator, $G_3(u)$ defined in
Eq.~(\ref{Gfreedman}), is a function of the scale-invariant variable
$u$, $h^{--}$ has scaling dimension $-2$ as it should.

Next we find the amplitude for particle 1 to propagate in this
background metric. This is just the bulk-to-bulk propagator $G(x, z;
x',z')$ for particle 1 in the presence of the shock wave at $x^+ = 0$
introduced by particle 2. Its equation is the same as for $G_5(u)$,
except for an additional term\footnote{Note that to linear order the inverse of
  $g_{MN} + h_{MN}$ is $g^{MN} - h^{MN}$, which implies that the
  shock potential in the propagator (\ref{eq:Gprop}) is $-h^{--}(x^+,x^\perp,z;x'^\perp,z')$ with the
  correct sign for gravitational attraction.}  for the contribution of $h^{--}$:
\be
[ 2 z^2\dd_{+} \dd_{-} - z^2 \dd^2_z + 3 z \dd_z  - z^2 \nabla^2_\perp 
+ R^2 h^{--}\dd^2_-] G(x,z;x',z') =  R^{5} \delta^4(x-x')\delta(z-z')/\sqrt{g} \; .
\label{eq:Gprop}
\ee
The metric $h^{--}(x^\perp,x^+, z)$
preserves translational invariance in $x^-$, so it is natural to
transform to fixed $p^+_1$,
\be
\widetilde G_{p^+_1}(x^+ - x'^+,x^\perp-x'^\perp,z,z') = \int dx^- e^{i p^+_1 (x^- - x'^-)} G(x,z;x',z') \ .
\ee
The resultant equation is just the light-cone Schroedinger equation
with ``time'' $\tau = x^+$ and conjugate ``Hamiltonian'' $H = P^-$:
\be
[ - i\dd_{\tau} + H] \widetilde G_{p^+_1}(x^+ - x'^+,x^\perp-x'^\perp,z,z')=   (z^3/2p^+_1) \delta^2(x^\perp-x'^\perp) \delta(x^+-x'^+)
\delta(z-z_0)
\ee
where
\be
2 p^+_1 H =  - \dd^2_z + 3 z^{-1} \dd_z  - \nabla^2_\perp 
-  (p^+_1)^2 R^2 z^{-2}  h^{--}(x^\perp, x^+, z) \; . 
\label{eq:H}
\ee
The solution is given by the time-ordered product 
\be
\widetilde G_{p^+_1}(x^+_3-x^+_1,x^\perp_3-x^\perp_1,z_3,z_1) =  \<x^\perp_3,z_3| \
T_\tau \left[\exp\left( -i \int^{x^+_3}_{x^+_1} d\tau H\right)\right] 
|x^\perp_1,z_1\>.
\ee
The Hamiltonian operator  has states enumerated by $|x^\perp,z\>$.
We can factorize this  into the product of three segments $\tau < 0, \tau = 0, \tau> 0$. 
\be
T_\tau \left[\exp\left( -i \int^{x^+_3}_{\epsilon} d\tau \hat H\right)\right]
 \exp\left(-i \int^{\epsilon}_{-\epsilon} d\tau H \right)
T_\tau \left[\exp\left( -i \int^{-\epsilon}_{x^+_1}d\tau \hat H\right)\right]
\ee
The first and the third factors above would contribute to the
bulk-to-boundary propagators, which are dropped when amputating the
Green's function.  The integral over $H$ for the middle term receives its
only contribution from the delta function in $ h^{--}(x^+) \sim  \chi \;
\delta(x^+) $, giving rise to the eikonal phase shift,
\be
 \chi(s,x^\perp-x'^\perp, z,z') = (\kappa_5^2/R^3) s z z' G_3(x^\perp-x'^\perp,z,z')/2 \; ,
\ee
for a (diagonal) unitary S matrix:
\be
S(s,x^\perp-x'^\perp, z,z') = e^{\textstyle i  \chi(s,x^\perp-x'^\perp, z,z')} 
\; .
\ee
This phase is in agreement with our earlier result for the truncated
bulk scattering amplitude,
\be
{\cal M}(s,x^\perp-x'^\perp, z,z')=  -2 is  \left(\frac{z z'}{R^2}\right)^2 \left [e^{\textstyle i \chi(s,x^\perp-x'^\perp, z,z')} \;  - \; 1\right] \;. 
\label{eq:norm}
\ee
In principle, one could derive the prefactor in this equation within
the context of the shock wave calculation; here we have merely matched
to the known tree-level amplitude.

\section{Conclusion}

We have considered the eikonal approximation to high-energy scattering
in the bulk of AdS space, as might be relevant for a portion of a
calculation of high-energy scattering in gauge theory, as well as
other physical processes.  We gave three approaches to the eikonal
amplitude: a heuristic picture for the $AdS$ scaling form, an explicit
resummation of Witten diagrams, and a shock wave derivation. All have
their advantages for further generalizations and clearer physical
intuition.

However, our results for the eikonal phase are valid only for
linearized semiclassical gravity.
For most physically important applications, the restrictions on our results
must be relaxed.  There are a number of technical and conceptual advances that
are needed, some of which are well within reach.
\begin{itemize}
\item Generalization to finite $\lambda$.  There is no obstruction to
extending our results to the case of Regge
behavior at finite $\lambda$, using \cite{BPST}.  Interesting
comparisons can be made with eikonal studies of flat-space string
theory, such as \cite{Amati:1993tb,Amati:1987wq,Amati:1992zb,Amati:1987uf}.
\item Generalization to non-conformal settings.  This
is necessary for a study of how the string theory realizes the
dual gauge theory's Froissart-Martin bound. In \cite{BPST} we
studied effects of confinement and running couplings, and again the
obstructions to extending our results to this case are purely technical.
\item Corrections to the eikonal approximation.
Full gauge-theory
computations require integrals over bulk coordinates $z$ and $z'$, but
the eikonal approximation is typically valid only in part of
the bulk; for instance, it may fail as $z\to z'$.
While reasonable approximations
will allow some gauge-theory computations to be carried out reliably, a
stronger understanding of scattering in all regions of
bulk coordinates is clearly desirable.  A minimal consistency requirement is that of ``local small angle" scattering, e.g., local momentum transfer should be less than the local energy. This has been commented upon briefly in the final section of \cite{BPST} and more extensively in  \cite{Brower:2007xg}.  In addition, one must also take into account other nonlinear effects.
\item Accounting for nonlinear corrections.  At small bulk impact
parameter, the gravitational fields of the scattering particles become
sufficiently large to require nonlinear gravity to be incorporated.
In some regimes, one must incorporate string interactions, such as
triple-Pomeron vertices; a role for effective Reggeon field theories
may be expected.  In other regimes one must include nonperturbative
effects, such as black holes
\cite{Veneziano:2004er,Eardley:2002re,Aharony:2005bm}.  Only in these
contexts can one begin to address questions of how field-theory
unitarity is restored at strong coupling, as relevant to high-energy
cross-sections, saturation phenomena, and heavy-ion collisions.  It is
clear that quantum string corrections must be addressed as well for
some relevant processes, if contact is to be made with QCD itself.
\end{itemize}

Despite the limited region of validity of our results, we see signs of
what we expect are general features that go well beyond this
regime. The eikonal phase is proportional to the Euclidean transverse
$AdS_3$ Green's function, a strong-coupling manifestation of conformal
symmetry of the gauge theory in the transverse plane, which is known
to arise for the weak-coupling BFKL kernel
\cite{Lipatov:1985uk,Kirschner:1989pw}.  The conventional picture in
4-d flat space, where the scattering particle picks up a phase at a
fixed position in the transverse impact parameter space $x^\perp=
(x_1,x_2)$, is generalized here to a phase at a fixed {\it bulk}
transverse position $x^\perp,z$.  In both the perturbative and shock
wave pictures, the exchange of an arbitrary number of rungs in a ladder
graph becomes effectively local, thus freezing
all transverse motion.  We expect these and other features will
survive, or be naturally extended, as other regimes are explored, and
a deeper understanding of high-energy scattering in gauge and string
theory emerges.

\

\noindent {\underline{Acknowledgments:}} 
We are pleased to acknowledge useful conversations
with D. Freedman, M. H. Fried, J. Polchinski, and G. Veneziano.  The
work of R.B. was supported by the Department of Energy under
Contract.~No.~DE-FG02-91ER40676, that of C.-I.T. was supported by the
Department of Energy under Contract~No.~DE-FG02-91ER40688, Task-A, and
that of M.J.S. by U.S. Department of Energy Contract.~No.~DE-FG02-96ER40956.  
C.-I.T. would like to thank the Aspen Center
for Physics for its hospitality during the writing of this paper.  We
are grateful to the Benasque Center for Science, where this work was
initiated.

\newpage
\bibliographystyle{utphys}
\bibliography{adseik_v2}

\providecommand{\href}[2]{#2}\begingroup\raggedright\begin{thebibliography}{10}

\bibitem{Veneziano:2004er}
G.~Veneziano, ``String-theoretic unitary s-matrix at the threshold of
  black-hole production,'' {\em JHEP} {\bf 11} (2004) 001,
\href{http://arXiv.org/abs/hep-th/0410166}{{\tt hep-th/0410166}}.

\bibitem{BPST}
R.~C. Brower, J.~Polchinski, M.~J. Strassler, and C.-I. Tan, ``The pomeron and
  gauge / string duality,''
\href{http://arXiv.org/abs/hep-th/0603115}{{\tt hep-th/0603115}}.

\bibitem{Cornalba:2006xk}
L.~Cornalba, M.~S. Costa, J.~Penedones, and R.~Schiappa, ``Eikonal
  approximation in ads/cft: From shock waves to four- point functions,''
\href{http://arXiv.org/abs/hep-th/0611122}{{\tt hep-th/0611122}}.

\bibitem{Cornalba:2006xm}
L.~Cornalba, M.~S. Costa, J.~Penedones, and R.~Schiappa, ``Eikonal
  approximation in ads/cft: Conformal partial waves and finite n four-point
  functions,'' {\em Nucl. Phys.} {\bf B767} (2007) 327--351,
\href{http://arXiv.org/abs/hep-th/0611123}{{\tt hep-th/0611123}}.

\bibitem{Giddings:2007bw}
S.~B. Giddings, D.~J. Gross, and A.~Maharana, ``Gravitational effects in
  ultrahigh-energy string scattering,''
\href{http://arXiv.org/abs/arXiv:0705.1816 [hep-th]}{{\tt arXiv:0705.1816
  [hep-th]}}.

\bibitem{Alday:2007hr}
L.~F. Alday and J.~Maldacena, ``Gluon scattering amplitudes at strong
  coupling,''
\href{http://arXiv.org/abs/arXiv:0705.0303 [hep-th]}{{\tt arXiv:0705.0303
  [hep-th]}}.

\bibitem{Cor}
L.~Cornalba, M.~S. Costa, and J.~Penedones, ``Eikonal approximation in ads/cft:
  Resumming the gravitational loop expansion,''
\href{http://arXiv.org/abs/arXiv:0707.0120 [hep-th]}{{\tt arXiv:0707.0120
  [hep-th]}}.

\bibitem{Polchinski:2001tt}
J.~Polchinski and M.~J. Strassler, ``Hard scattering and gauge/string
  duality,'' {\em Phys. Rev. Lett.} {\bf 88} (2002) 031601,
\href{http://arXiv.org/abs/hep-th/0109174}{{\tt hep-th/0109174}}.

\bibitem{DIS}
J.~Polchinski and M.~J. Strassler, ``Deep inelastic scattering and gauge/string
  duality,'' {\em JHEP} {\bf 05} (2003) 012,
\href{http://arXiv.org/abs/hep-th/0209211}{{\tt hep-th/0209211}}.

\bibitem{Brower:2002er}
R.~C. Brower and C.-I. Tan, ``{Hard scattering in the M-theory dual for the QCD
  string},'' {\em Nucl. Phys.} {\bf B662} (2003) 393--405,
\href{http://arXiv.org/abs/hep-th/0207144}{{\tt hep-th/0207144}}.

\bibitem{Freedman}
E.~D'Hoker, D.~Z. Freedman, S.~D. Mathur, A.~Matusis, and L.~Rastelli,
  ``Graviton exchange and complete 4-point functions in the ads/cft
  correspondence,'' {\em Nucl. Phys.} {\bf B562} (1999) 353--394,
\href{http://arXiv.org/abs/hep-th/9903196}{{\tt hep-th/9903196}}.

\bibitem{BFKL1}
L.~N. Lipatov, ``Reggeization of the vector meson and the vacuum singularity in
  nonabelian gauge theories,'' {\em Sov. J. Nucl. Phys.} {\bf 23} (1976)
338--345.

\bibitem{BFKL2}
E.~A. Kuraev, L.~N. Lipatov, and V.~S. Fadin, ``The pomeranchuk singularity in
  nonabelian gauge theories,'' {\em Sov. Phys. JETP} {\bf 45} (1977)
199--204.

\bibitem{BFKL3}
I.~I. Balitsky and L.~N. Lipatov, ``The pomeranchuk singularity in quantum
  chromodynamics,'' {\em Sov. J. Nucl. Phys.} {\bf 28} (1978)
822--829.

\bibitem{CW}
H.~Cheng and T.~T. Wu, ``Impact factor and exponentiation in high-energy
  scattering processes,'' {\em Phys. Rev.} {\bf D186} (1969)
1611--1618.

\bibitem{Kabat:1992tb}
D.~Kabat and M.~Ortiz, ``Eikonal quantum gravity and planckian scattering,''
  {\em Nucl. Phys.} {\bf B388} (1992) 570--592,
\href{http://arXiv.org/abs/hep-th/9203082}{{\tt hep-th/9203082}}.

\bibitem{Giudice:2001ce}
G.~F. Giudice, R.~Rattazzi, and J.~D. Wells, ``Transplanckian collisions at the
  lhc and beyond,'' {\em Nucl. Phys.} {\bf B630} (2002) 293--325,
\href{http://arXiv.org/abs/hep-ph/0112161}{{\tt hep-ph/0112161}}.

\bibitem{Amati:1993tb}
D.~Amati, M.~Ciafaloni, and G.~Veneziano, ``Effective action and all order
  gravitational eikonal at planckian energies,'' {\em Nucl. Phys.} {\bf B403}
  (1993)
707--724.

\bibitem{'tHooft:1987rb}
G.~'t~Hooft, ``Graviton dominance in ultrahigh-energy scattering,'' {\em Phys.
  Lett.} {\bf B198} (1987)
61--63.

\bibitem{'tHooft:1990fr}
G.~'t~Hooft, ``The black hole interpretation of string theory,'' {\em Nucl.
  Phys.} {\bf B335} (1990)
138--154.

\bibitem{D'Hoker:1999jc}
E.~D'Hoker, D.~Z. Freedman, S.~D. Mathur, A.~Matusis, and L.~Rastelli,
  ``Graviton and gauge boson propagators in ads(d+1),'' {\em Nucl. Phys.} {\bf
  B562} (1999) 330--352,
\href{http://arXiv.org/abs/hep-th/9902042}{{\tt hep-th/9902042}}.

\bibitem{Chang:1971je}
S.-J. Chang and T.-M. Yan, ``High-energy elastic and inelastic scattering in
  phi-to-the- third theory,'' {\em Phys. Rev.} {\bf D4} (1971)
537--558.

\bibitem{Levy:1969}
M.~Levy and J.~Sucher {\em Phys. Rev.} {\bf D186} (1969)
1665--.

\bibitem{Amati:1987wq}
D.~Amati, M.~Ciafaloni, and G.~Veneziano, ``Superstring collisions at planckian
  energies,'' {\em Phys. Lett.} {\bf B197} (1987)
81.

\bibitem{Fried:1990}
H.~M. Fried, ``{``Basics of Functional Methods and Eikonal Models."
  Gif-sure-Yvette, France: Ed. Frontieres (1990)},''.

\bibitem{Fried:2002ds}
H.~M. Fried and Y.~Gabellini, ``Summing all the eikonal graphs. ii,'' {\em Eur.
  Phys. J.} {\bf C32} (2003) 55--65,
\href{http://arXiv.org/abs/hep-th/0208057}{{\tt hep-th/0208057}}.

\bibitem{Amati:1992zb}
D.~Amati, M.~Ciafaloni, and G.~Veneziano, ``Planckian scattering beyond the
  semiclassical approximation,'' {\em Phys. Lett.} {\bf B289} (1992)
87--91.

\bibitem{Kabat:1992pz}
D.~Kabat, ``Validity of the eikonal approximation,'' {\em Comments Nucl. Part.
  Phys.} {\bf 20} (1992) 325--336,
\href{http://arXiv.org/abs/hep-th/9204103}{{\tt hep-th/9204103}}.

\bibitem{Aichelburg:1970dh}
P.~C. Aichelburg and R.~U. Sexl, ``On the gravitational field of a massless
  particle,'' {\em Gen. Rel. Grav.} {\bf 2} (1971)
303--312.

\bibitem{Arcioni:2001my}
G.~Arcioni, S.~de~Haro, and M.~O'Loughlin, ``Boundary description of planckian
  scattering in curved spacetimes,'' {\em JHEP} {\bf 07} (2001) 035,
\href{http://arXiv.org/abs/hep-th/0104039}{{\tt hep-th/0104039}}.

\bibitem{Kang:2004jd}
K.~Kang and H.~Nastase, ``High energy {QCD} from planckian scattering in {AdS}
  and the froissart bound,'' {\em Phys. Rev.} {\bf D72} (2005) 106003,
\href{http://arXiv.org/abs/hep-th/0410173}{{\tt hep-th/0410173}}.

\bibitem{Kang:2005bj}
K.~Kang and H.~Nastase, ``Heisenberg saturation of the froissart bound from
  {AdS-CFT},'' {\em Phys. Lett.} {\bf B624} (2005) 125--134,
\href{http://arXiv.org/abs/hep-th/0501038}{{\tt hep-th/0501038}}.

\bibitem{Amati:1987uf}
D.~Amati, M.~Ciafaloni, and G.~Veneziano, ``Classical and quantum gravity
  effects from planckian energy superstring collisions,'' {\em Int. J. Mod.
  Phys.} {\bf A3} (1988)
1615--1661.

\bibitem{Brower:2007xg}
R.~C. Brower, M.~J. Strassler, and C.-I. Tan, ``{On The Pomeron at Large 't
  Hooft Coupling},''
\href{http://arXiv.org/abs/0710.4378}{{\tt 0710.4378}}.

\bibitem{Eardley:2002re}
D.~M. Eardley and S.~B. Giddings, ``Classical black hole production in
  high-energy collisions,'' {\em Phys. Rev.} {\bf D66} (2002) 044011,
\href{http://arXiv.org/abs/gr-qc/0201034}{{\tt gr-qc/0201034}}.

\bibitem{Aharony:2005bm}
O.~Aharony, S.~Minwalla, and T.~Wiseman, ``Plasma-balls in large n gauge
  theories and localized black holes,'' {\em Class. Quant. Grav.} {\bf 23}
  (2006) 2171--2210,
\href{http://arXiv.org/abs/hep-th/0507219}{{\tt hep-th/0507219}}.

\bibitem{Lipatov:1985uk}
L.~N. Lipatov, ``The bare pomeron in quantum chromodynamics,'' {\em Sov. Phys.
  JETP} {\bf 63} (1986)
904--912.

\bibitem{Kirschner:1989pw}
R.~Kirschner and L.~N. Lipatov, ``Bare reggeons in asymptotically free
  theories,'' {\em Z. Phys.} {\bf C45} (1990)
477.

\end{thebibliography}\endgroup

\end{document}